\documentclass[11pt]{article}
\usepackage{amssymb}
\usepackage{amsfonts}
\usepackage{latexsym}
\usepackage{subfigure}
\usepackage{graphicx}
\usepackage{amsmath}
\usepackage{cite}
\usepackage{wrapfig}
\usepackage{caption}
\usepackage{float}
\usepackage{breqn}
\usepackage[pdfpagelabels=true,plainpages=false,colorlinks=true,linkcolor=blue,citecolor=blue,urlcolor=blue]{hyperref}
\usepackage{subfigure}
\makeatletter
\newcommand{\doublespacing}{\let\CS=\@currsize\renewcommand{\baselinestretch}{1.35}\tiny\CS}

\newtheorem{theorem}{Theorem}

\begin{document}
\newcommand{\be}{\begin{equation}}
\newcommand{\ee}{\end{equation}}
\newcommand{\bea}{\begin{eqnarray}}
\newcommand{\eea}{\end{eqnarray}}
\newcommand{\nn}{\nonumber}
\newcommand{\ba}{\begin{array}}
\newcommand{\ea}{\end{array}}
\newcommand{\lb}{\label}
\newcommand{\ben}{\begin{enumerate}}
\newcommand{\een}{\end{enumerate}}
\newcounter{saveeqn}
\newcommand{\alpheqn}{\setcounter{saveeqn}{\value{equation}}%
\stepcounter{saveeqn}\setcounter{equation}{0}%
\renewcommand{\theequation}{\mbox{\arabic{saveeqn}\alph{equation}}}}
\newcommand{\reseteqn}{\setcounter{equation}{\value{saveeqn}}%
\renewcommand{\theequation}{\arabic{equation}}}
%\pagestyle{myheadings}
%\markboth{}
%\bibliographystyle{elsarticle-num}
\thispagestyle{empty}
\title{ \bf  {Forced convection in a fluid saturated anisotropic porous channel with isoflux boundaries}}
\author {\bf Timir Karmakar$^{1}$,~Motahar Reza$^{2}$,~G. P. Raja Sekhar$^{1}$
\\
Department of Mathematics\\ Indian Institute of Technology Kharagpur,
Kharagpur - 721302, India\\
$^{2}$ School of Computer Science and Engineering,\\ National Institute of Science and Technology, Berhampur, India\\
}
\date{}
\maketitle
\vspace{-0.3in }
{\bf Abstract}\\
Fully developed forced convective flow in a channel filled with porous material bounded by two impermeable walls subject to constant heat flux is considered. We consider Brinkman-Forchhimer equation to govern the flow inside the porous medium which accounts for the presence of inertial term. We assume that the porous medium is anisotropic in nature and permeability is varying along all the directions, correspondingly, permeability will be a matrix which appear as a positive semi-definite matrix in the momentum equation. The anisotropic nature of the porous medium is characterized by anisotropic ratio and angle. We have obtained the velocity, temperature and Nusselt number numerically, due to the presence of the non-linear quadratic term in the momentum equation. The effect of anisotropic permeability ratio and anisotropic angle on hydrodynamic and heat transfer is shown. We see that Nusselt number is altered significantly with the anisotropic ratio and angle. We present a fresh analysis about the inclusion of the permeability matrix in the Brinkman-Forchhimer extended Darcy momentum equation. To the best of the authors knowledge, this is the first attempt to analyze general anisotropy corresponding to Brinkman-Forchhimer extended Darcy model.

%{\bf Keywords}~~Squeezing force, Anisotropic permeability ratio, Anisotropic angle, Convection-diffusion-reaction, Asymptotic method.

\section{Introduction}
Forced convection in porous media has drawn the attention of many researchers due to its diverse application in geothermal systems, thermal insulations, solid matrix heat exchangers, nuclear waste disposal etc. \cite{Nield1996,Nield2003,Nield2004,Olek1998}. A simple and widely used model in this regard is fully developed convective flow in a porous channel. A comprehensive idea pertaining to this can be found in the research monograph by Nield and Bejan \cite{Nield-book}. The existing literature regarding this mostly focused to Darcy and Brinkman models. However, incorporation of inertia effects in the momentum equation make the problem difficult to obtain the velocity field explicitly \cite{Hooman2006}. Inclusion of inertia effects in the momentum equation results a non-linear differential equation due to a non-linear term (Forchheimer term \cite{Nield-book}) which accounts for the viscous drag along with the Brinkman equation. The corresponding equation very often is referred as Brinkman-Forchheimer equation in literature \cite{Vafai1981,Vafai1989}. From the physical perspective, quadratic term which appears in the momentum equation is an inclusion of inertial drag force which is comparable for large filtration velocity and cannot be neglected.\cite{Nield-book}

There are significant attempts to analyze the forced convection inside porous ducts using Brinkman-Forchhimer model. Vafai and Tien \cite{Vafai1981} analyzed the problem considering inertia, and solved by applying a boundary layer approximation. In their analysis they gave specific attention to the flow in the proximity of the solid boundary. Kaviany \cite{Kaviany1985} studied laminar flow in a porous channel bounded by isothermal plates, though neglected the inertial term for computational difficulty. It is shown that Nusselt number may significantly increase with increasing the porous media shape parameter. Poulikakos and Renken \cite{Poulikakos1987} studied forced convection in a channel including the effect of flow inertia, variable porosity and Brinkamn friction. Renken and Poulikakos \cite{Renken1988} presented an experimental investigation with numerical analysis of forced convection in a packed bed of spheres. They reported that overall heat transfer is enhanced with the inclusion of inertial term along with the Darcy term. Vafai and Kim \cite{Vafai1989} presented an exact solution of forced convection in a porous channel with inertial effect based on boundary layer approximation. The corresponding analysis presented was not uniformly valid for all values of fluid flow and porous media properties, such as, viscosity, inertia, permeability, etc. However, this work gives a comprehensive discussion about the dependency of various flow characteristics such as velocity, temperature, Nusselt number, on inertial parameter, which motivates the authors to do the present work. Nield et al. \cite{Nield1996} reinvestigated the problem presented by Vafai and Kim \cite{Vafai1989} and presented an analytical solution which is valid uniformly for all combination of parameters, corresponding to constant heat flux and constant temperature at the wall. They pointed the issues those are flawed in the analysis by Vafai and Kim \cite{Vafai1989} and presented a comprehensive discussion about heat transfer coefficient dependency on inertial coefficient and viscosity ratio. More studies concerning flow and heat transfer involving porous media accounting inertial effects can be found in \cite{Huang2010,Lage1996,Kuznetsov1996,Hooman2006,Nield-book}.

\par

Available literature displays that a number of experimental and theoretical investigations have been done to understand forced convection in a porous duct considering the inertia effects corresponding to isotropic porous situation. However, the corresponding problem when the porous medium is anisotropic in nature is less explored. Some literatures involving flow and heat transfer can be found in Rees et al. \cite{Rees2002}, Hill and Morad \cite{Hill2014}, Qin and Kaloni \cite{Qin1994}, Starughan and Walker \cite{Straughan1996}, Payne and Straughan \cite{Payne1998l}, Rees and Postelnicu \cite{Rees2001}. Porous materials those are used in heat transfer devices, such as porous foam, collection of tubes, fibrous insulation materials, are anisotropic in nature (see Boomsma et al. \cite{Boomsma2003}, Nakayama et al. \cite{Nakayama_anisotropy2002}, Kim et al. \cite{Kim_anisotropy2001}). The arbitrary orientation of pores or grains forming the porous medium cause anisotropy (see Rees et al. \cite{Rees2002}, Hill and Morad \cite{Hill2014}, Riley and Rees \cite{Riley1985non}, Karmakar and Raja Sekhar \cite{Karmakar2017}). Very often a pack of tubes is used in heat transfer devices to have efficient heat transfer. Such a collection of tubes is an example of an anisotropic porous medium (see Yang and Lee \cite{Yang1999}, Lee and Yang \cite{Lee_anisotropy1998}). The fluid flow within the gaps of the tube depends on the orientation of the tubes. For isotropic porous medium the permeability matrix is constant, while the same is treated as a tensor for anisotropic porous medium. Permeabilities vary with direction for anisotropic porous medium. This anisotropic permeability tensor in the momentum equation makes the governing equations become less user friendly to treat analytically.
%=====================================================================================
\par
Modern day industrial needs involving hydro-mechanical systems work with the aid of fluid flow and heat transfer . In order to have efficient heat transfer, it is common to use porous packings \cite{London1949,Liebenberg2007,Lage_Patent1999}. Also it is a usual practice to use dampers to avoid the flow speed and optimize convective heat transfer. Heat transfer efficiency also can be increased by disrupting the fluid boundary layer for high speed flows (see, Delavar et al. \cite{Delavar2013}, Huang et al. \cite{Huang2010}). Changing the physical and materialistic properties of the medium in which fluid flows, for example, flow inside anisotropic porous media may also alter the heat transfer rate (see, Nakayama et al. \cite{Nakayama_anisotropy2002}, Kim et al. \cite{Kim_anisotropy2001}). With the inclusion of anisotropic permeability and considering the inertia effect, the forced convection inside porous medium is very limited. Kim et al. \cite{Kim_anisotropy2001} studied the effect of anisotropy on thermal performance of an aluminium foam sink while considering the quadratic Forchimmer term in the momentum equation. However, their study was restricted to permeability varies only along the principle axis. Nakayama et al. \cite{Nakayama_anisotropy2002} determined the Forchimmer drag with the inclusion of anisotropy. Though these few studies analyze the effect of anisotropy in the presence of inertia, however, the idea is very subtle. In view of this we expect that considering the permeability variations along all the direction (introducing anisotropic angle; see, Karmakar and Raja Sekhar \cite{Timir2016a}), and analyzing the effect of anisotropy on forced convection in the presence of inertia will be interesting. Furthermore, this analysis provides more physical insights to understand the fluid flow and heat transfer inside more complex structured porous medium.

\begin{figure}[h]
\centering
\includegraphics[height= 6.2 cm, width=10 cm]{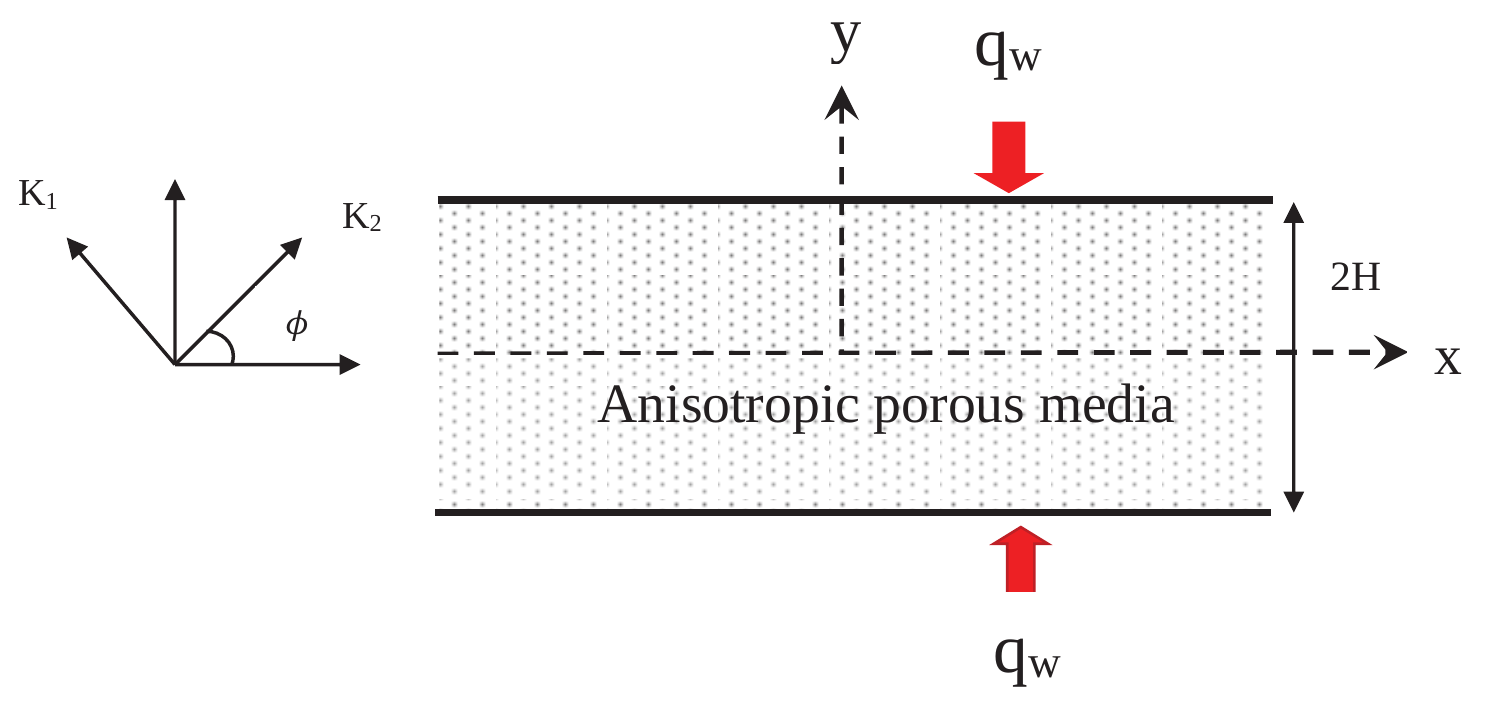}
\caption{Schematic diagram of the physical situation}\label{fig:geomtery}
\end{figure}

%As far as the authors knowledge, none of them considered inertial effects along with viscous dissipation while considering the porous medium as anisotropic in nature \cite{Timir2016a,Timir_dissipation,Degan2002forced}. The corresponding problem considering anisotropic porous media is less explored. However, in many cases especially from industry, porous material posses an anisotropic structure. For example fluid flow and heat transfer through a collection of cylindrical tube is very often posses an anisotropic network due to its orientation (see Yang and Lee \cite{Yang1999}). Considering anisotropic porous media can provide more information about the fluid flow and heat transfer in the corresponding medium. This gives impetus to attempt the present study.

We consider a fully developed forced convection in a channel filled with anisotropic porous medium. The permeability of the medium is varying along all the direction in view of an angle (referred as anisotropic angle in literature). In section \ref{Mathematical formulation} we present the problem description and mathematical formulation. We then proceed to find numerical solution and results in section \ref{Numerical Results}. We see that anisotropy of the porous medium has significant effect on both the hydrodynamics and heat transfer witnessing the presence of inertial term.

\section{Mathematical formulation}\label{Mathematical formulation}
The problem of interest is shown in Fig. \ref{fig:geomtery}. We consider a rectangular channel filled with a porous material which is bounded by two impermeable walls at $\widetilde{y}=\pm H$. The flow is assumed to be steady, along $\widetilde{x}$-direction and fully developed. The permebilties along two principal axes is $K_{1}$ and $K_{2}$, those are constant. The orientation angle very often referred as anisotropic angle is defined as the angle between the horizontal direction and the principal axis with permeability $K_{2}$. Before considering the fully developed flow in horizontal direction, we consider a generic form of the corresponding mass conservation and momentum equation. Correspondingly, the mass conservation and the momentum equation are given by \cite{Nield-book}
\begin{equation}\label{eq1}
\nabla.\widetilde{\textbf{u}}=0,
\end{equation}
\begin{equation}\label{eq2}
-\nabla \widetilde{p}+ \mu_{eff} \nabla^{2} \widetilde{\textbf{u}}-\mu \textbf{K}^{-1} \widetilde{\textbf{u}}-c_{F} \rho_{f} \textbf{K}^{-1/2} |\widetilde{\textbf{u}}|\widetilde{\textbf{u}}=0,
\end{equation}
where $\mu_{eff}$ is the effective viscosity of the fluid inside the porous medium, $\mu$ is the dynamic viscosity and $c_{F}$ is the inertial coefficient. Here $\textbf{K}$ corresponds to the permeability of the porous medium. Since the porous medium is anisotropic in nature the permeability is a second order tensor given by \cite{Timir2016a,Timir_dissipation,Degan2002forced}
\begin{equation} \label{eq3}
\textbf{K}=\left(
             \begin{array}{cc}
               K_{1}\sin^{2} \phi+K_{2}\cos^{2} \phi & (K_{2}-K_{1})\sin \phi \cos \phi \\
               (K_{2}-K_{1})\sin \phi \cos \phi & K_{2} \sin^{2} \phi+K_{1}\cos^{2} \phi \\
             \end{array}
           \right).
\end{equation}

\begin{theorem}\label{theorem1}
\cite{Horn2012matrix} Let $A\in M_{n}$ be Hermitian and positive semidefinite matrix and let $k\in\left\{2,3,4...\right\}$, then there exist a unique Hermitian positive semidefinite matrix $B$ such that $B^{k}=A$.
\end{theorem}
Clearly the above matrix $\mathbf{K}$ is symmetric and its eigenvalues are $K_{1}$ and $K_{2}$. Since $K_{1},K_{2}\geq 0$, hence, $\mathbf{K}$ is positive semidefinite. By Theorem \ref{theorem1} one may get a unique positive semidefinite matrix $B$ such that $B^{k}=\mathbf{K}$. Hence $\mathbf{K}^{-1/2}$ exist and it is semi positive definite and unique.
\par

The assumption of fully developed flow along $x$-direction gives $\widetilde{\textbf{u}}=(\widetilde{u}(\widetilde{y}),0)$. Under this assumptions the conservation of mass and momentum reduces to
\begin{equation}\label{eq4}
\frac{\partial \widetilde{u}}{\partial \widetilde{x}}=0,
\end{equation}
\begin{equation}\label{eq5}
-\frac{\partial \widetilde{p}}{\partial \widetilde{x}}+ \mu_{eff} \frac{d^{2}\widetilde{u}}{d\widetilde{y}^{2}}-\mu \left(\frac{K_{2}\sin^{2}\phi+K_{1}\cos^{2}\phi}{K_{1}K_{2}}\right)\widetilde{u}-
\rho_{f}c_{F}\left(\frac{\sqrt{K_{1}} \cos^{2}\phi+\sqrt{K_{2}}\sin^{2}\phi}{\sqrt{K_{1}K_{2}}}\right)\widetilde{u}^{2}=0,
\end{equation}
\begin{equation}\label{eq6}
-\frac{\partial \widetilde{p}}{\partial \widetilde{y}}-\frac{(K_{1}-K_{2})\sin\phi \cos\phi}{K_{1}K_{2}}\widetilde{u}-\rho_{f}c_{F}\frac{(\sqrt{K_{1}}-\sqrt{K_{2}})\sin\phi \cos\phi}{\sqrt{K_{1}K_{2}}}\widetilde{u}^{2}=0.
\end{equation}
with the boundary conditions
\begin{equation}\label{bcs1}
\widetilde{u}=0 \quad \textrm{at} \quad \widetilde{y}=H,
\end{equation}
\begin{equation}\label{bcs2}
\frac{d\widetilde{u}}{d\widetilde{y}}=0 \quad \textrm{at} \quad \widetilde{y}=0.
\end{equation}
Eq. (\ref{eq5}) and Eq. (\ref{eq6}) together imply $\partial \widetilde{p}/\partial \widetilde{x}=\textrm{const}=-G~(\textrm{say})$.
Hence the one-dimensional momentum equation takes the form
\begin{equation}\label{final_eq}
-\frac{\partial \widetilde{p}}{\partial \widetilde{x}}+ \mu_{eff} \frac{d^{2}\widetilde{u}}{d\widetilde{y}^{2}}- \frac{\mu a }{K_{1}}\widetilde{u}- \frac{\rho_{f}c_{F} b}{\sqrt{K_{1}}}\widetilde{u}^{2}=0,
\end{equation}
where
$a=\sin^{2}\phi+ K \cos^{2}\phi$, $b=\sin^{2}\phi+\sqrt{K} \cos^{2}\phi$ are the parameters characterizing the anisotropy and $K=K_{1}/K_{2}$ is the anisotropic permeability ratio.

We introduce the following non-dimensional variables
\begin{equation}
y=\frac{\widetilde{y}}{H},~ u=\frac{\mu \widetilde{u}}{G H^{2}}, ~M=\frac{\mu_{eff}}{\mu},~ Da=\frac{K_{1}}{H^{2}}, ~F=\frac{c_{F}\rho_{f}G H^{3}}{\mu^{2}},
\end{equation}
where $M$ is the viscosity ratio, $Da$ is the Darcy number, $F$ is the Forchheimer number.
After the non-dimensionalization Eq. (\ref{final_eq}) takes the form
\begin{equation}\label{momentum_eqn}
M \frac{d^{2}u}{dy^{2}}- \epsilon^{2} a u-F b \epsilon u^{2}+1=0,
\end{equation}
where $\epsilon=1/\sqrt{Da}$ is the porous media shape parameter.

The corresponding boundary conditions after non-dimensionalization reduce to
\begin{equation}\label{boundary1}
u=0 \quad \textrm{at} \quad y=1,
\end{equation}
\begin{equation}\label{boundary2}
\frac{du}{dy}=0 \quad \textrm{at} \quad y=0.
\end{equation}
The bulk mean velocity (in dimensional form) of the system is given by
\begin{equation}
\overline{\widetilde{u}}=\frac{1}{H} \int_{0}^{H} \widetilde{u} d\widetilde{y}.
\end{equation}
Corresponding non-dimensional mean velocity is given by
\begin{equation}
\overline{u}=\int_{0}^{1} u dy.
\end{equation}

Assuming the homogeneity and local thermal equilibrium, the steady-state energy equation for the current configuration can be written as \cite{Hung2008,Hooman2008}
\begin{equation}\label{energy_equation}
\rho C_{p} u \frac{\partial \widetilde{T}}{\partial \widetilde{x}}=k \frac{\partial^{2}\widetilde{T}}{\partial \widetilde{y}^{2}},
\end{equation}
where $T$ is the local temperature, $k$ is the thermal conductivity, $C_{p}$ is the specific heat.

The assumption of constant heat flux at the wall and symmetry condition leads to
\begin{equation}
\frac{\partial \widetilde{T}}{\partial \widetilde{y}}=\frac{q_{w}}{k} \quad \textrm{at} \quad \widetilde{y}=H \quad \textrm{and} \quad \frac{\partial \widetilde{T}}{\partial \widetilde{y}}=0 \quad \textrm{at} \quad \widetilde{y}=0.
\end{equation}
The bulk mean temperature of the system is given by
\begin{equation}\label{mean_temperature}
T_{m}=\frac{1}{H\overline{\widetilde{u}}}\int_{0}^{H} \widetilde{u }\widetilde{T} d\widetilde{y}.
\end{equation}
Also it follows from the first law of thermodynamics that \cite{Nield1996,Nield-book}
\begin{equation}
\frac{\partial \widetilde{T}}{\partial \widetilde{x}}=\frac{q_{w}}{\rho C_{p}H\overline{\widetilde{u}}}.
\end{equation}

We introduce further non-dimensional variables and re-normalize as
\begin{equation}\label{re_non_dim}
\Theta=\frac{\widetilde{T}-\widetilde{T}_{w}}{\widetilde{T}_{m}-\widetilde{T}_{w}}, \quad U=\frac{u}{\overline{u}}, \quad Nu=\frac{2Hq_{w}}{k(\widetilde{T}_{w}-\widetilde{T}_{m})}
\end{equation}
where $Nu$ is the Nusselt number which characterizes the heat transfer between the wall and the porous medium, $Br$ is Brinkman number which corresponds to the viscous dissipation.

 Assuming thermally fully developed flow along axial direction and scaling the variables by following Eq. (\ref{re_non_dim}), the energy equation (\ref{energy_equation}) reduces to \cite{Nield1996,Nield-book,Hooman2004}
\begin{equation}\label{energy_dissipation}
2 \frac{d^{2}\Theta}{dy^{2}}+ Nu U= 0,
\end{equation}
which is to be solved subject to
\begin{equation}\label{boundary3}
\Theta(1)=0 \quad \textrm{and} \quad \frac{d\Theta}{dy}(0)=0.
\end{equation}
Nusselt number can be found using the compatibility condition
\begin{equation}\label{compatibility1}
\int_{0}^{1} \Theta U dy=1.
\end{equation}

\section{Numerical results and discussion} \label{Numerical Results}
The momentum equation given in Eq. (\ref{momentum_eqn}) is a seconde order non-linear differential equation which is solved using fourth order Runge-Kutta method with the help of Shooting technique using the boundary condition given in Eqs. (\ref{boundary1}) and (\ref{boundary2}). To compute the Nusselt number we assume the temperature of the form $\Theta(y)=Nu \widehat{\theta}(y)$, where the expression of $\widehat{\theta}(y)$ can be obtained from Eq. (\ref{energy_dissipation}) as
\begin{equation}\label{numerical1}
\widehat{\theta}(y)=\frac{1}{2} \int_{y}^{1} \int_{0}^{\xi_{2}} U(\xi_{1}) d\xi_{1} d\xi_{2}.
\end{equation}
We use Simpson's fourth order method to evaluate the double integration given in Eq. (\ref{numerical1}) with number of intervals $1000$ for higher accuracy. Using the obtained numerical value of $\widehat{\theta}(y)$ and $U(y)$, we finally obtain the Nusselt number using the compatibility condition (\ref{compatibility1}) as
\begin{equation}\label{compatability2}
Nu=\frac{1}{\int_{0}^{1}\widehat{\theta}(y)U(y)}.
\end{equation}

\subsection{Hydrodynamic analysis}
\begin{figure}[h]
\centering
\subfigure[]{\label{Fig_2}\includegraphics[height= 6 cm, width= 7 cm]{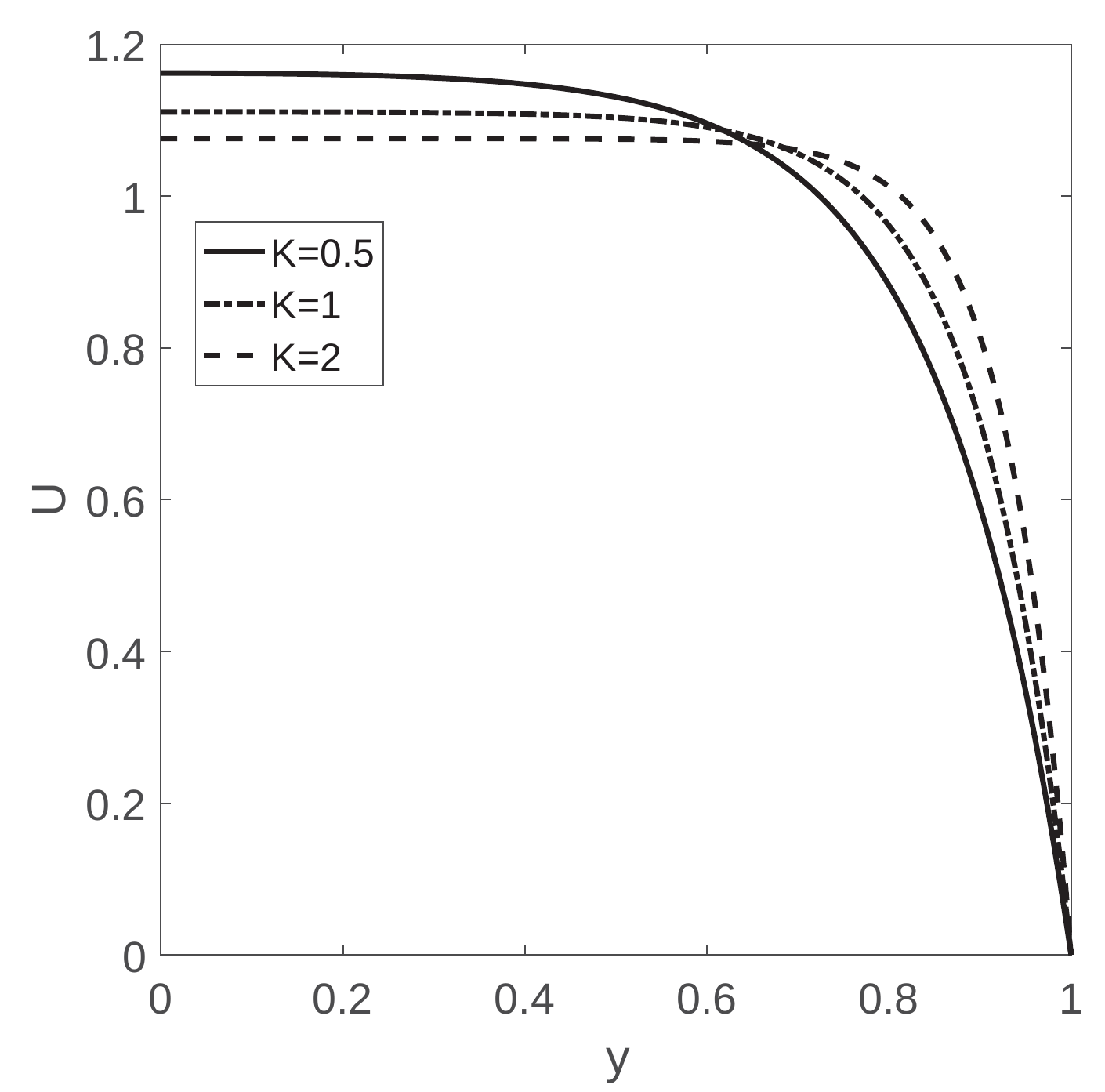}}
\subfigure[]{\label{Fig_3}\includegraphics[height= 6 cm, width= 7 cm]{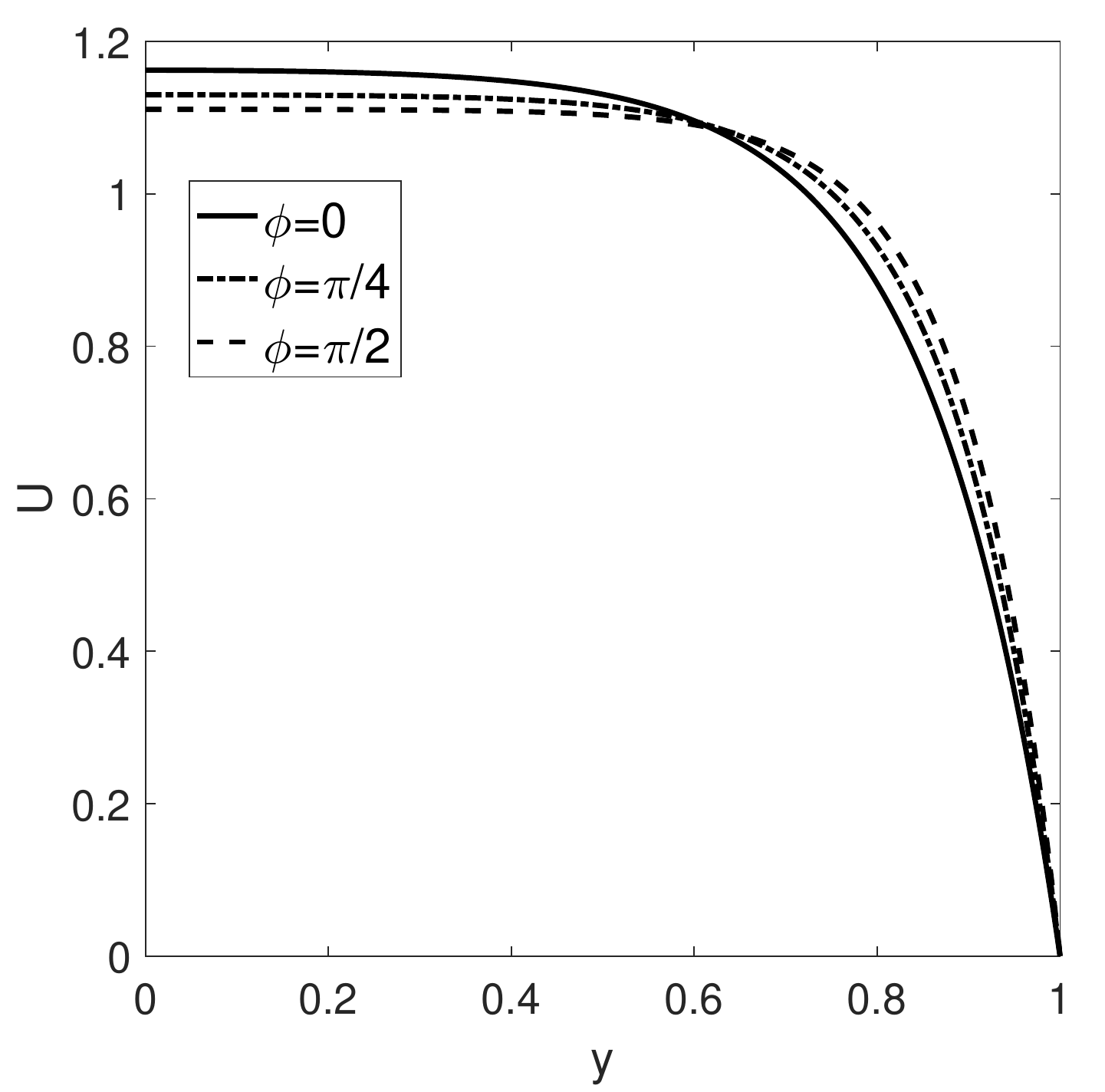}}
\caption{(a) Velocity profile for different anisotropic ratio with $Da=10^{-2}$, $\phi=0$, $M=1$, $F=1$; (b) Velocity profile for different anisotropic angle with $Da=10^{-2}$, $K=0.5$, $M=1$, $F=1$.}
\end{figure}

Fig. \ref{Fig_2} depicts the velocity profile for different anisotropic permeability ratio, $K$. We see that for a fixed value of $Da(<1)$ with increasing anisotropic ratio $K$, the magnitude of the velocity decreases. This is as expected, because for a fixed $Da$, i.e., $K_{1}$, increasing in permeability ratio $K=K_{1}/K_{2}$ results reduced permeability in the horizontal direction, i.e., the flow direction, as $\phi=0$. Hence the magnitude of velocity decreases. Fig. \ref{Fig_3} depicts the velocity profile for different anisotropic angle, $\phi$. We see that when $K<1$, the magnitude of the velocity is maximum for $\phi=0$ and minimum for $\phi=\pi/2$. This behavior is in conformity with the fact that for $K<1$ for a fixed $Da$, i.e., $K_{1}$, $\phi=0$ results enhanced permeability in the horizontal direction (flow direction). However, this behavior is opposite when $K>1$, i.e., the magnitude of the velocity is maximum for $\phi=\pi/2$ and minimum for $\phi=0$ (results not presented here).

\begin{figure}[h]
\centering
\subfigure[]{\label{Fig_4}\includegraphics[height= 6 cm, width= 7 cm]{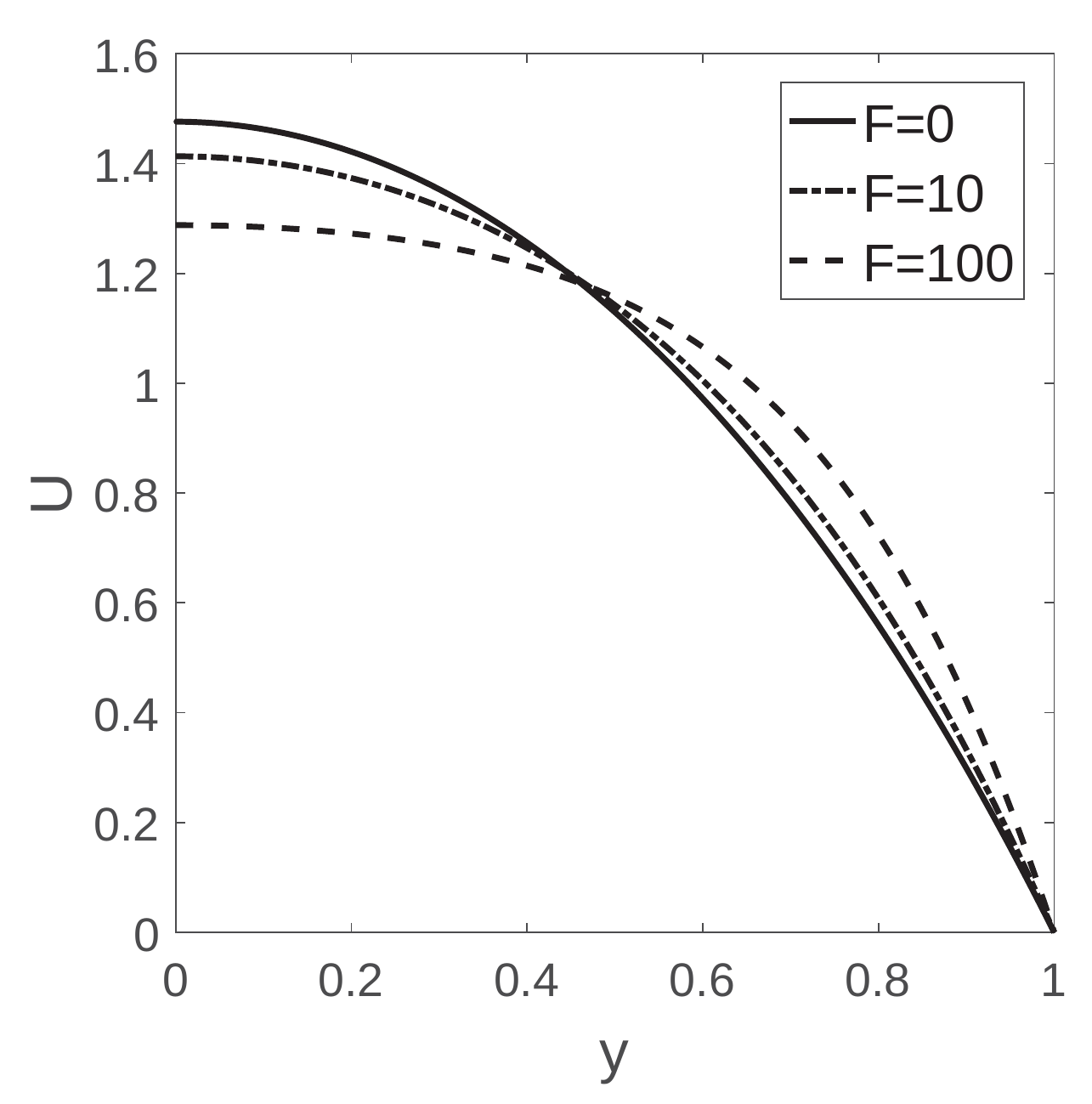}}
\subfigure[]{\label{Fig_5}\includegraphics[height= 6 cm, width= 7 cm]{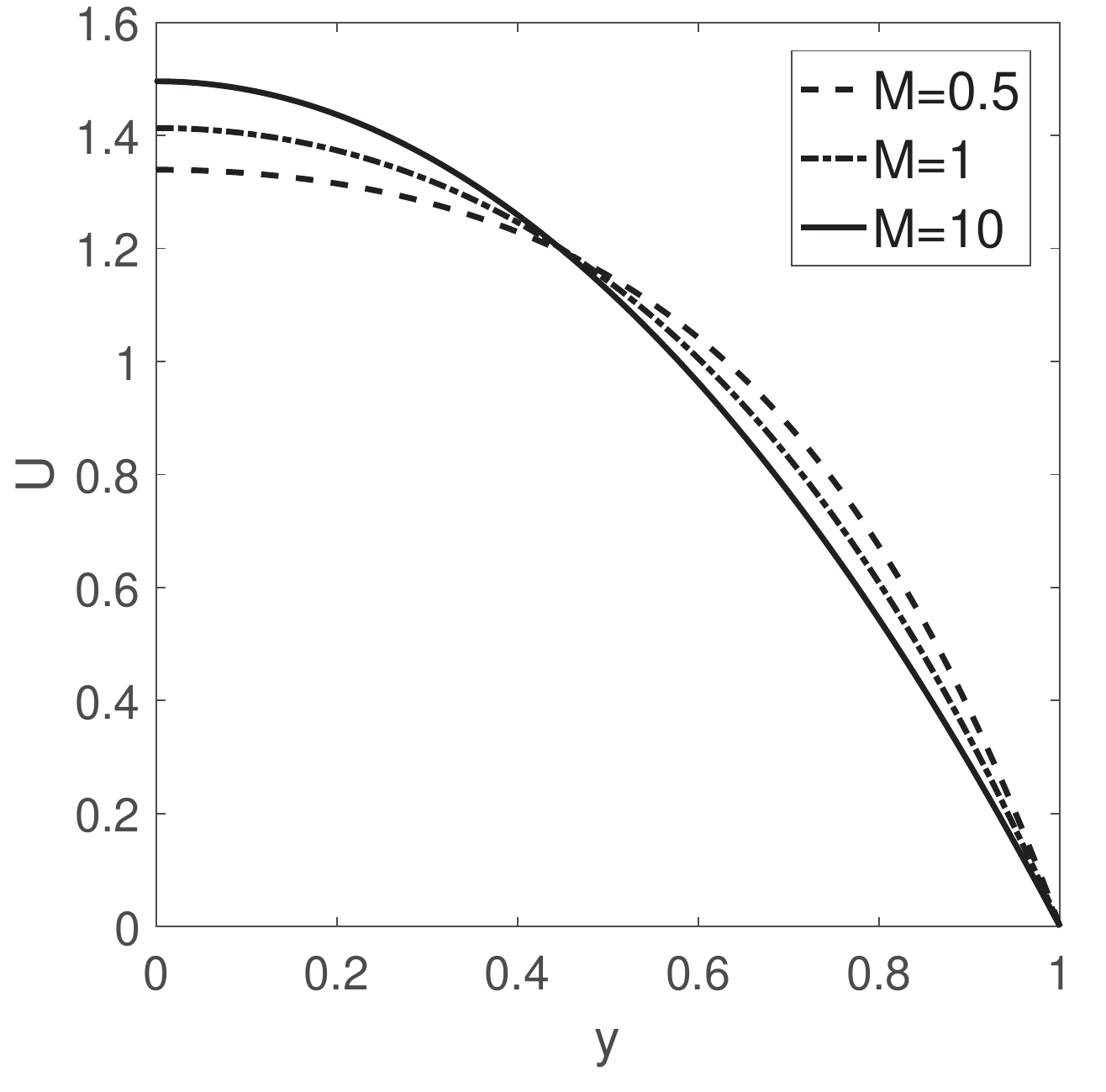}}
\caption{(a) Velocity for  different Forchheimer number ($F$) for $Da=1$, $\phi=0$, $M=1$, $K=1$; (b) Velocity for  different viscosity ratio ($M$) for $Da=1$, $\phi=0$, $F=1$, $K=1$.}
\end{figure}

The effect of variation of $F$ is shown in Fig. \ref{Fig_4} for fixed $Da$, $M$ and $K$. We see that the magnitude of the velocity decreases with $F$. The velocity is scaled by a physical quantity which is proportional to the applied pressure gradient. With increasing $F$ the asymptotic behavior of velocity shows that it is proportional to the square root of the applied pressure gradient, in other words, the resistance to the flow increases. One may observe a similar discussion in Nield et al. \cite{Nield1996}. Hence, the velocity is flattened near the center of the channel with increasing $F$. The scenario is opposite near the channel wall due to a constant volumetric flow rate throughout the channel. We see that viscosity ratio $M$ has significant effect on the velocity (see Fig. \ref{Fig_5}). We see that as the viscosity ratio increases the magnitude of the velocity increases. An order of magnitude analysis on Eq. (\ref{momentum_eqn}) reveals that the measure of the thickness of the momentum boundary layer is $\left(M Da/a\right)^{1/2}$. For a fixed $M$, the velocity is maximum at the center and decreases towards the wall to retain a constant volumetric flux. However, with increasing $M$, the distortion of the velocity profile is less near the wall and show larger distortion at the center. We have normalized the non-dimensional velocity with mean velocity and hence this constant volumetric flux constraint adjusts the flow with n the width of the channel and hence we notice a cross over at a particular width $y$ (see Fig. \ref{Fig_5}).

% Hence, we see that the distortion of the velocity profile is decreasing with increasing $M$ near the channel wall and increasing with increasing $M$ near the center, due to a constant volumetric flow rate. The calculated results depicted in Fig. \ref{Fig_5} are agreeing with this fact. The results depicted in Nield et al. \cite{Nield1996} show an decreasing behavior of the velocity with increasing $M$ throughout the channel cross-section. The results in the present work is relatively different due to the plotted normalized velocity $U(y)$, which is obtained by scaling $u(y)$ by the mean velocity $\overline{u}$. One can get a similar variation of viscosity ratio $M$ on the velocity as Nield et al. \cite{Nield1996}, if $u(y)$ is plotted instead of $U(y)$.

\subsection{Heat transfer analysis}

\begin{figure}[h]
\centering
\subfigure[]{\label{Fig_6}\includegraphics[height= 6.3 cm, width= 7.3 cm]{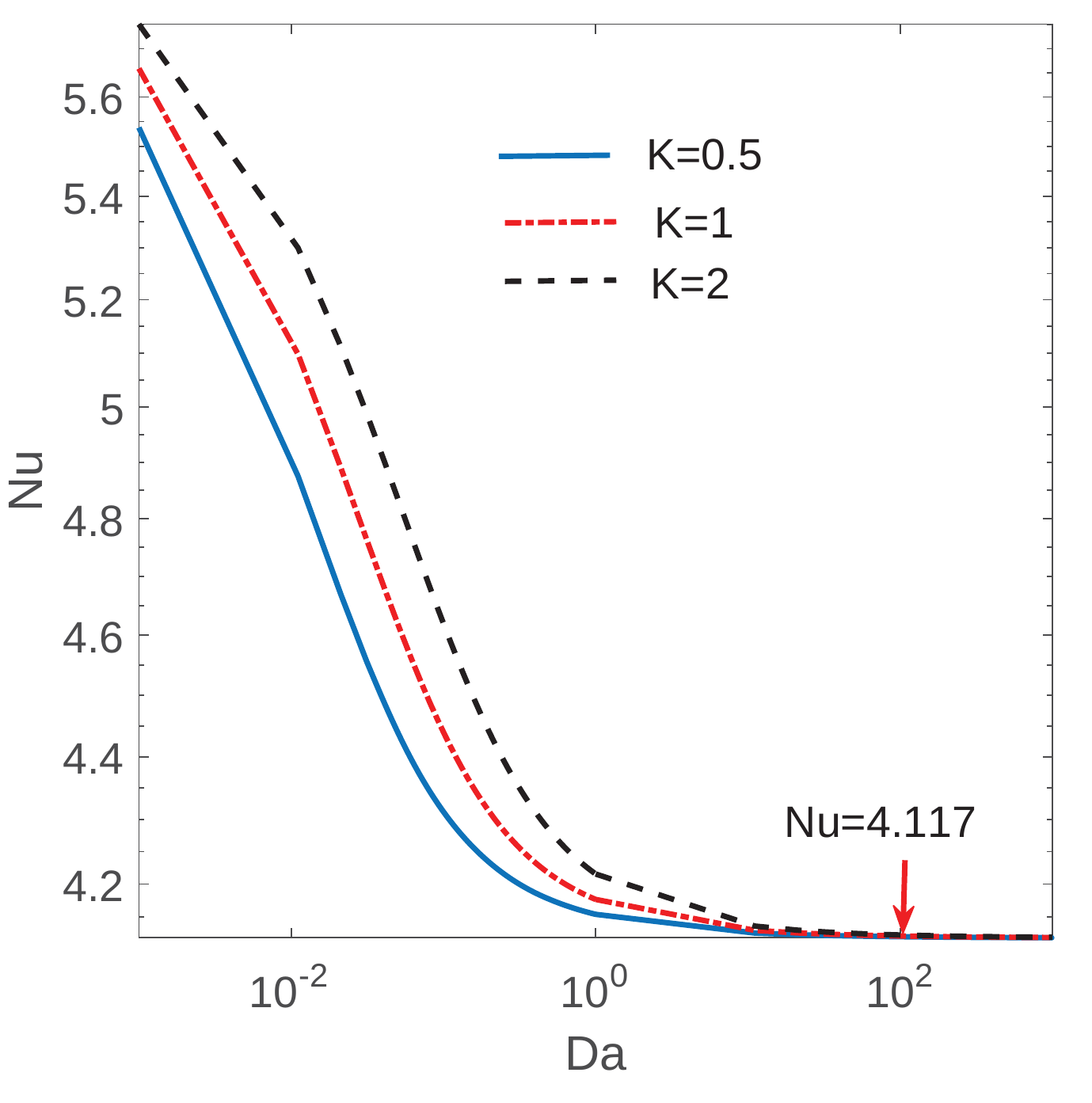}}
\subfigure[]{\label{Fig_7}\includegraphics[height= 6.3 cm, width= 7.3 cm]{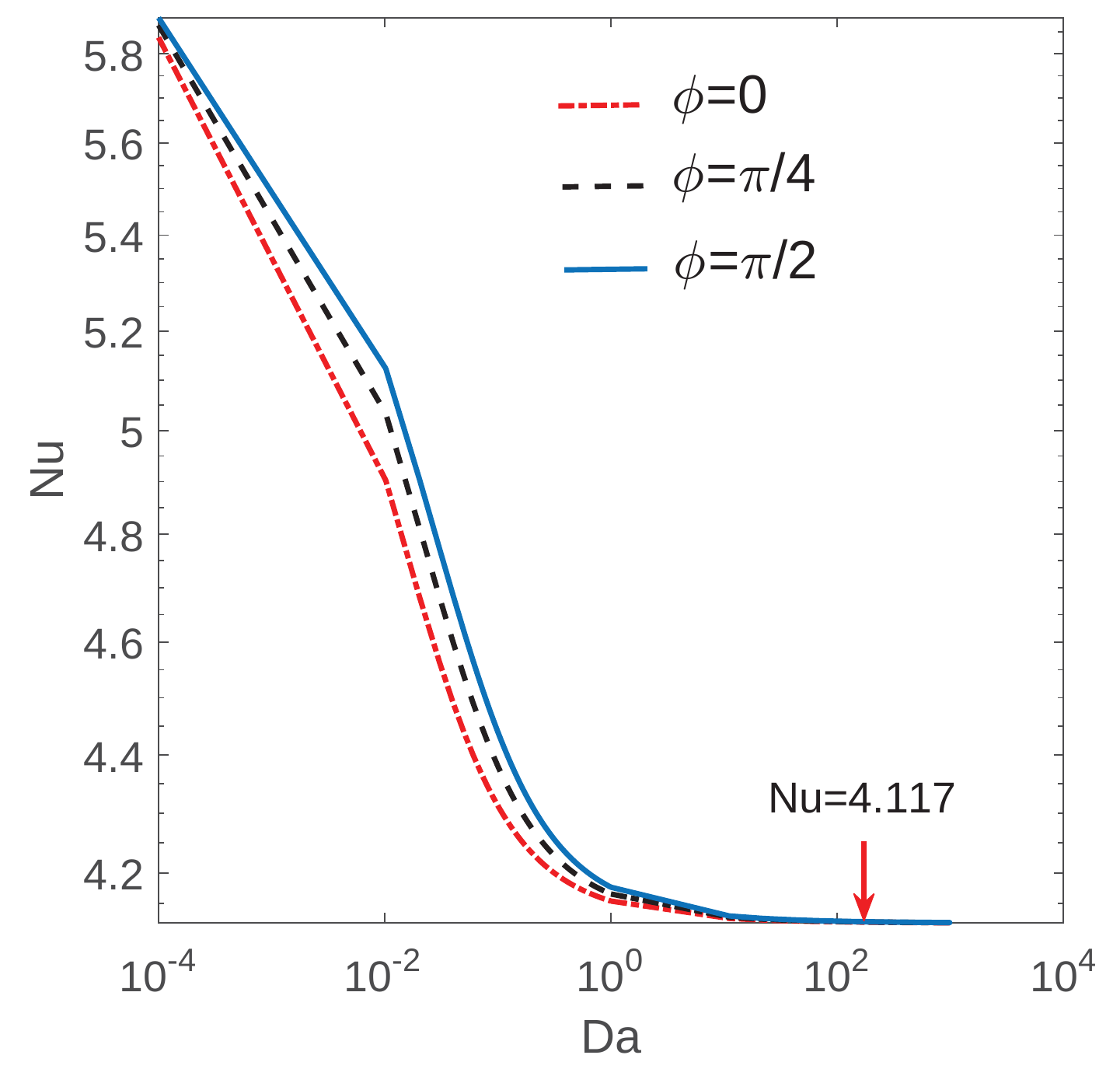}}
\caption{Effect of Darcy number, $Da$, on the Nusselt number (a) for different anisotropic ratio, $K$, for $F=1$, $M=1$, $\phi=0$; (b) for different anisotropic angle, $\phi$, for $F=1$, $M=1$, $K=0.5$.}
\end{figure}

We have shown the Nusselt number, $Nu$, variation as a function of $Da$ for different anisotropic ratio. We see that $Nu$ approaches to the value $4.117$ for large values of $Da$ and close to $6$ for low $Da$. We see that the anisotropic permeability ratio has significant effect for moderate values of Darcy number. The increasing value of $Nu$ with the anisotropic ratio can be explained if we look at the definition of $Nu$ given in Eq. (\ref{compatability2}). As the anisotropic ratio increases the fluid velocity decreases and tends to more uniform (see Fig. \ref{Fig_2}). As a result the denominator term of the Eq. (\ref{compatability2}) decreases and hence, $Nu$ increases. The mean temperature is defined in Eq. (\ref{mean_temperature}) is weighted by the velocity. When the velocity is less, tends to a slug flow situation then the mean temperature $T_{m}$ is just the usual ordinary temperature. Hence, with increasing anisotropic ratio the flow becomes uniform in the middle and velocity decreases near the channel wall due to viscous effect. As a result the difference $\left(T_{w}-T_{m}\right)$ decreases and correspondingly Nusselt number, $Nu$, increases. Figure \ref{Fig_7} shows the Nusselt number variation as a function of $Da$ for different anisotropic angle, $\phi$. We see that the Nusselt number show larger deviation in the moderate Darcy regime. Nusselt number decreases as the anisotropic angle increases, and maximum when $\phi=\pi/2$. We can draw a similar discussion since the velocity is minimum for $\phi=\pi/2$ when $K<1$ (see Fig. \ref{Fig_3}). However, we do not repeat here.

\begin{figure}[H]
\centering
\subfigure[]{\label{Fig_8}\includegraphics[height= 6 cm, width= 7 cm]{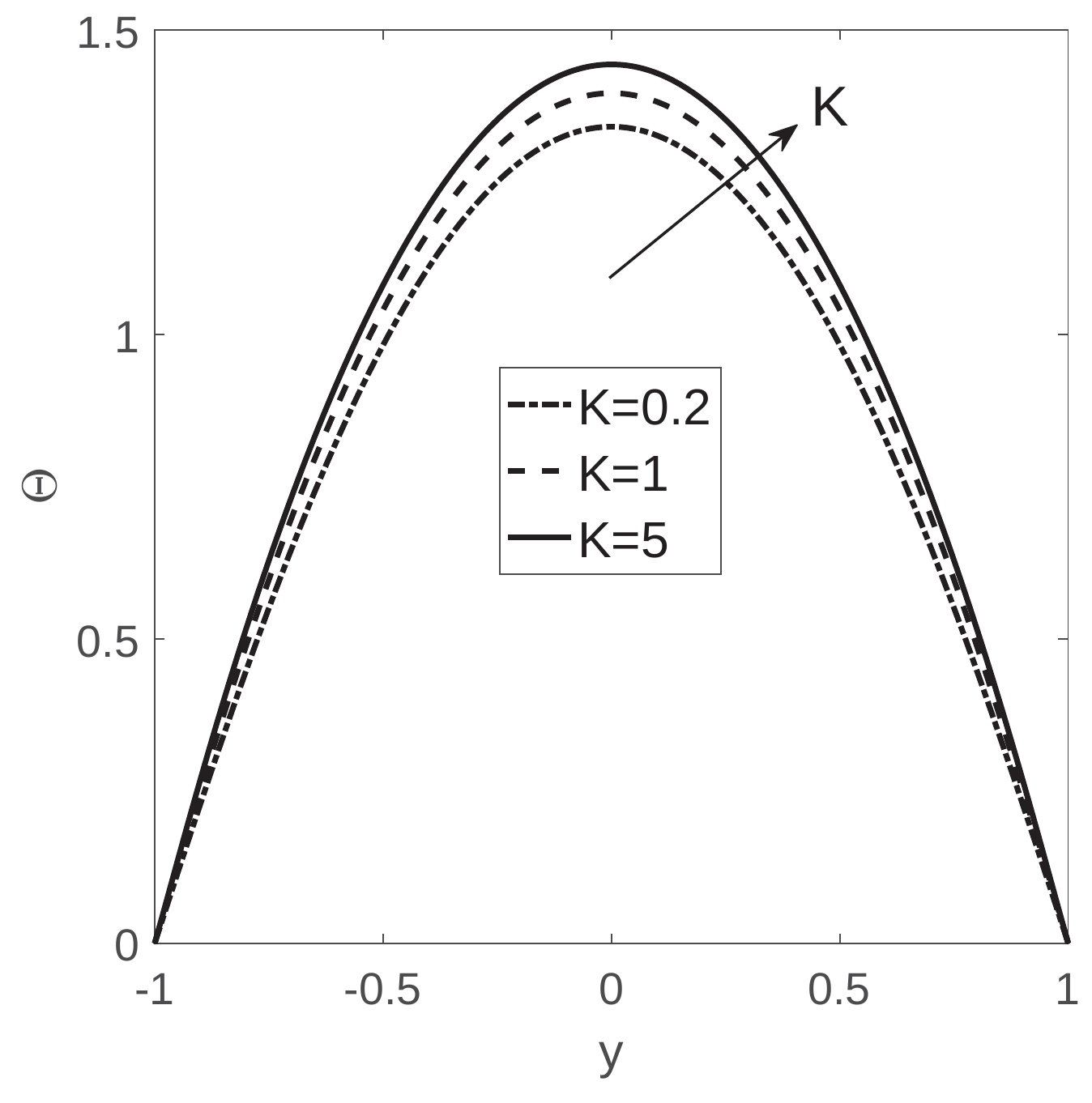}}
\subfigure[]{\label{Fig_9}\includegraphics[height= 6 cm, width= 7 cm]{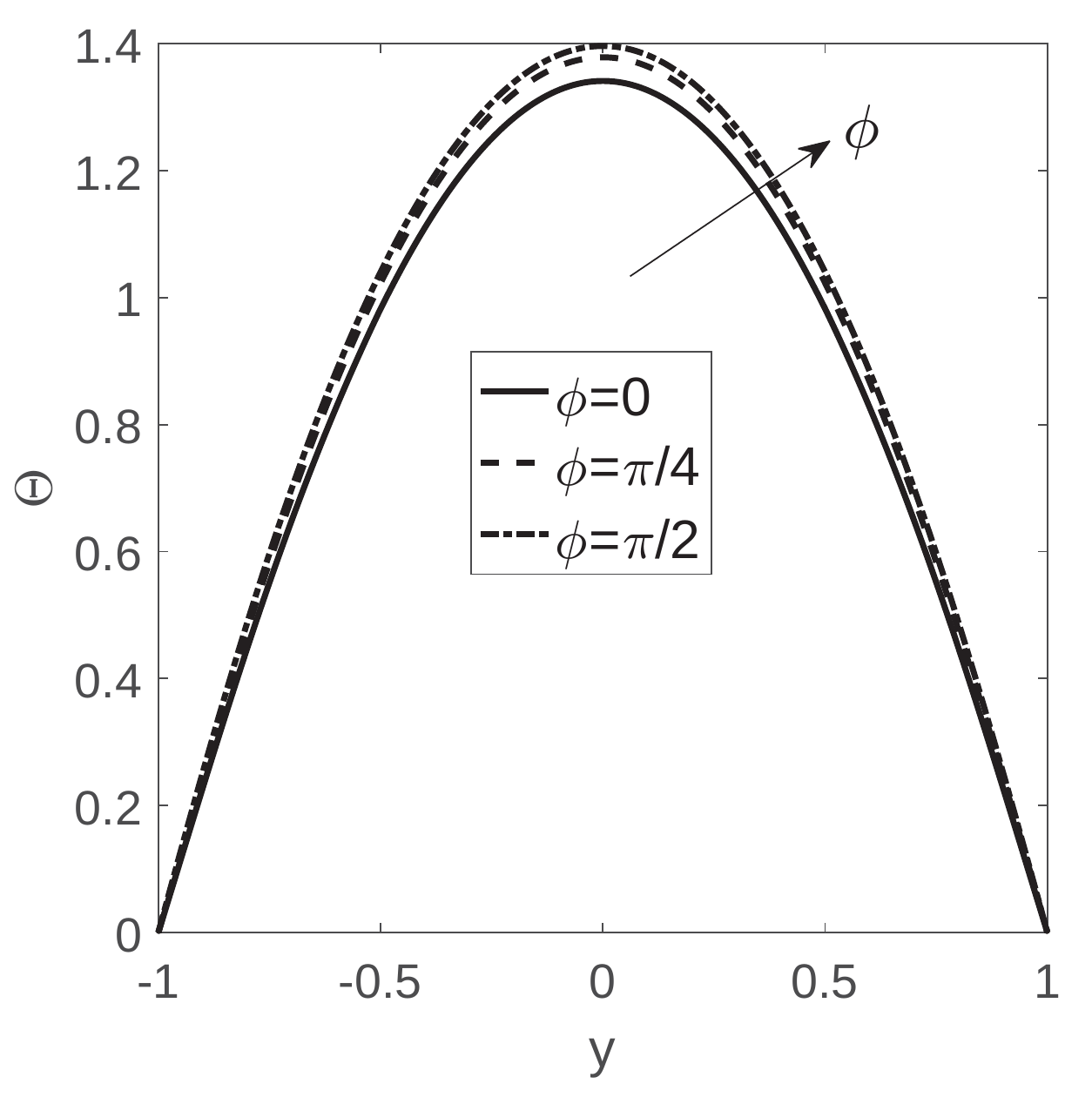}}
\caption{(a) Temperature profile for different anisotropic ratio for $Da=10^{-2}$, $\phi=0$, $M=1$, $F=1$; (b) Temperature profile for different anisotropic angle for $Da=10^{-2}$, $K=0.2$, $F=1$, $M=1$.}
\end{figure}

Figure \ref{Fig_8} shows the temperature distribution for different anisotropic permeability ratio, $K$. We see that increasing $K$ leads to increasing non-dimensional temperature $\Theta$. This is conformal with the fact that, as the anisotropic ration increases $Nu$ increases, causes enhanced heat transfer and hence the temperature. A similar type of discussion prevails for the temperature distribution for various anisotropic angle, $\phi$. As we indicated earlier, for $\phi=\pi/2$, Nusselt number is high, so the non-dimensional temperature $\Theta$ is high (see Fig. \ref{Fig_9}).

\begin{figure}[h]
\centering
\subfigure[]{\label{Fig_10}\includegraphics[height= 6 cm, width= 7 cm]{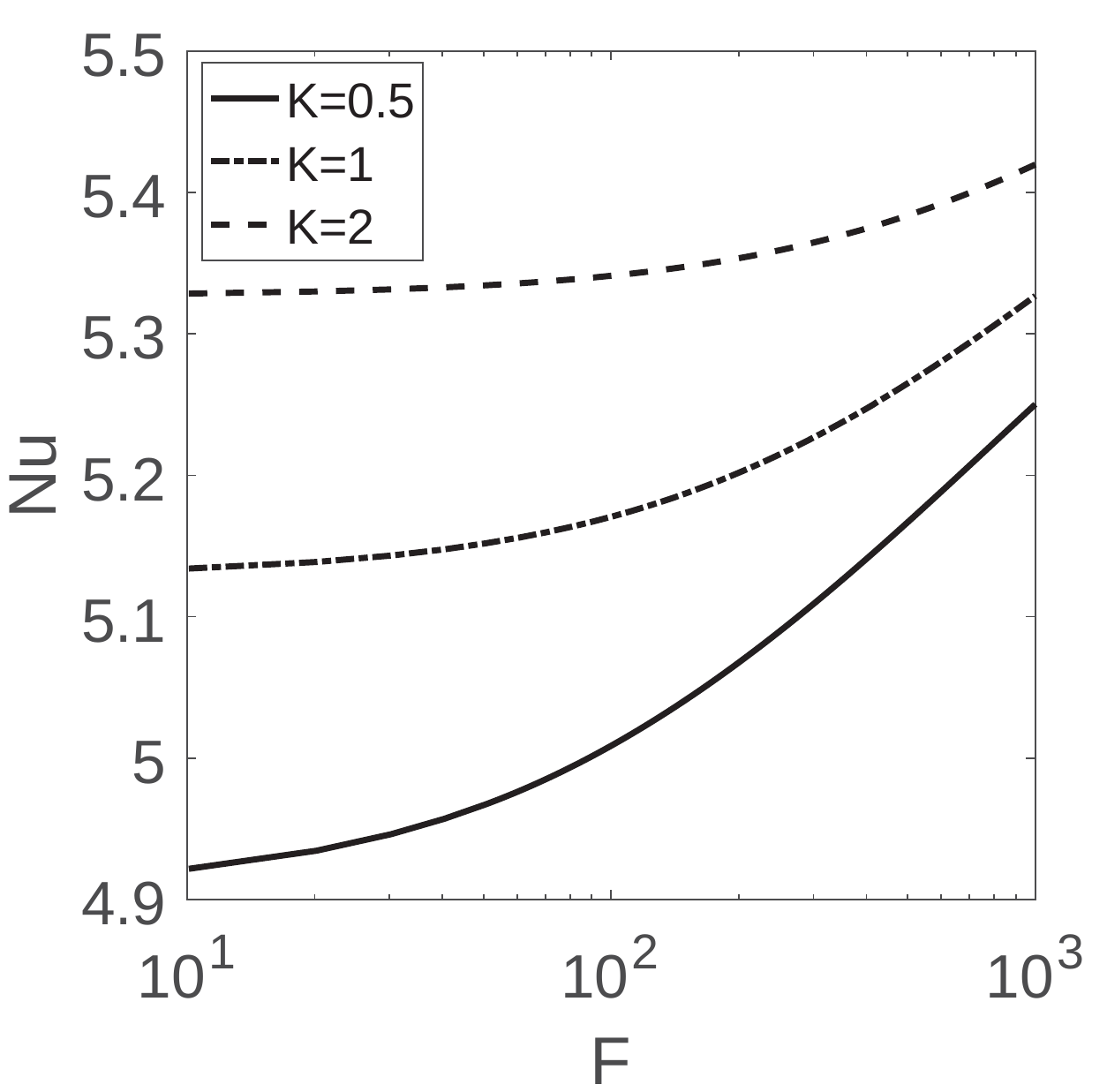}}
\subfigure[]{\label{Fig_11}\includegraphics[height= 6 cm, width= 7 cm]{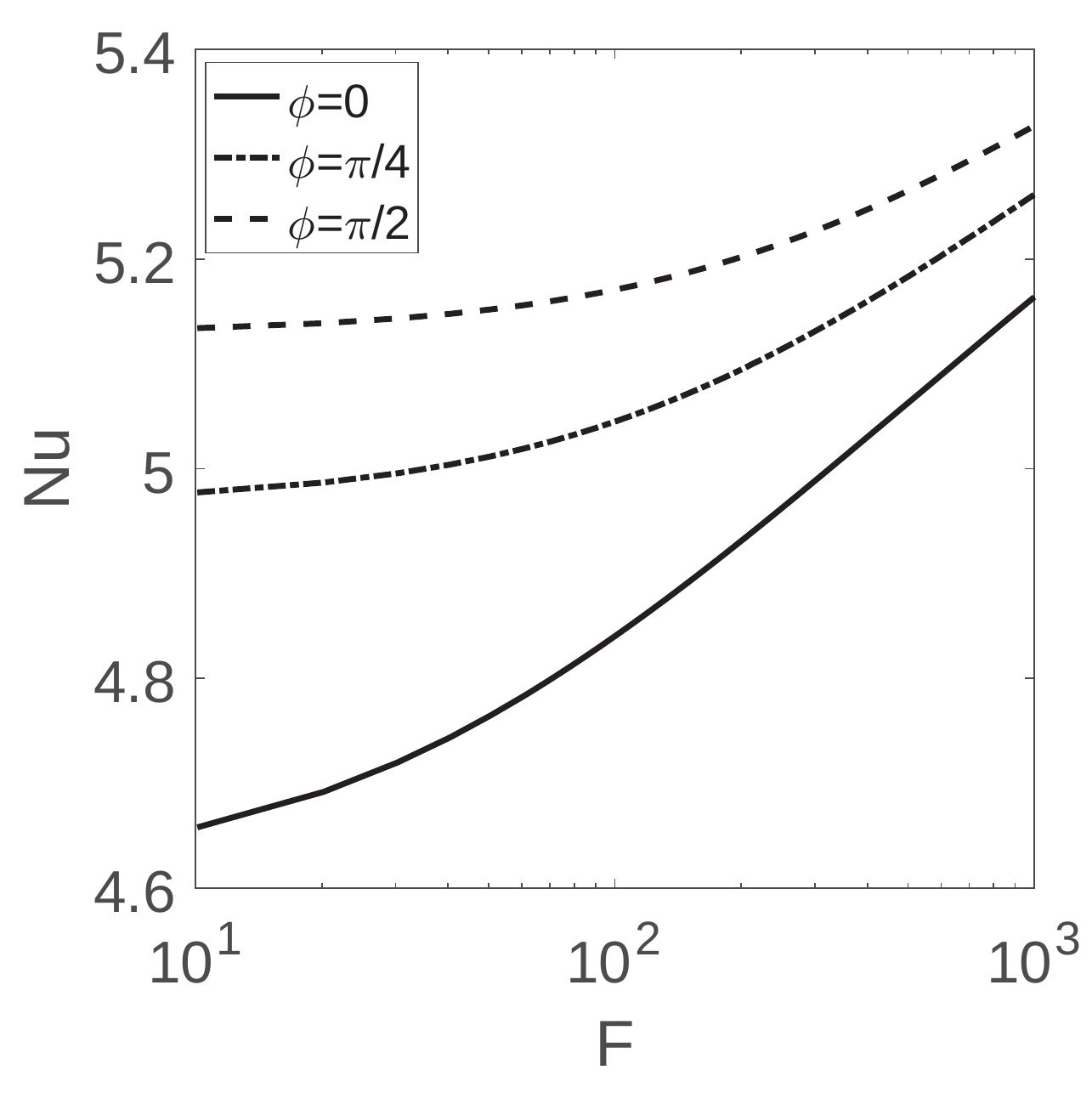}}
\caption{Effect of Forchheimer number, $F$, on the Nusselt number (a) for different anisotropic ratio, $K$, for $Da=10^{-2}$, $M=1$, $\phi=0$; (b) for different anisotropic angle, $\phi$,  for $Da=10^{-2}$, $M=1$, $K=0.2$; (c) for different viscosity ratio, $M$, for $Da=10^{-2}$, $K=1$.}\label{Fig_6_all}
\end{figure}

\begin{figure}[h]
\centering
\subfigure[]{\label{Fig_1*}\includegraphics[height= 6 cm, width= 7 cm]{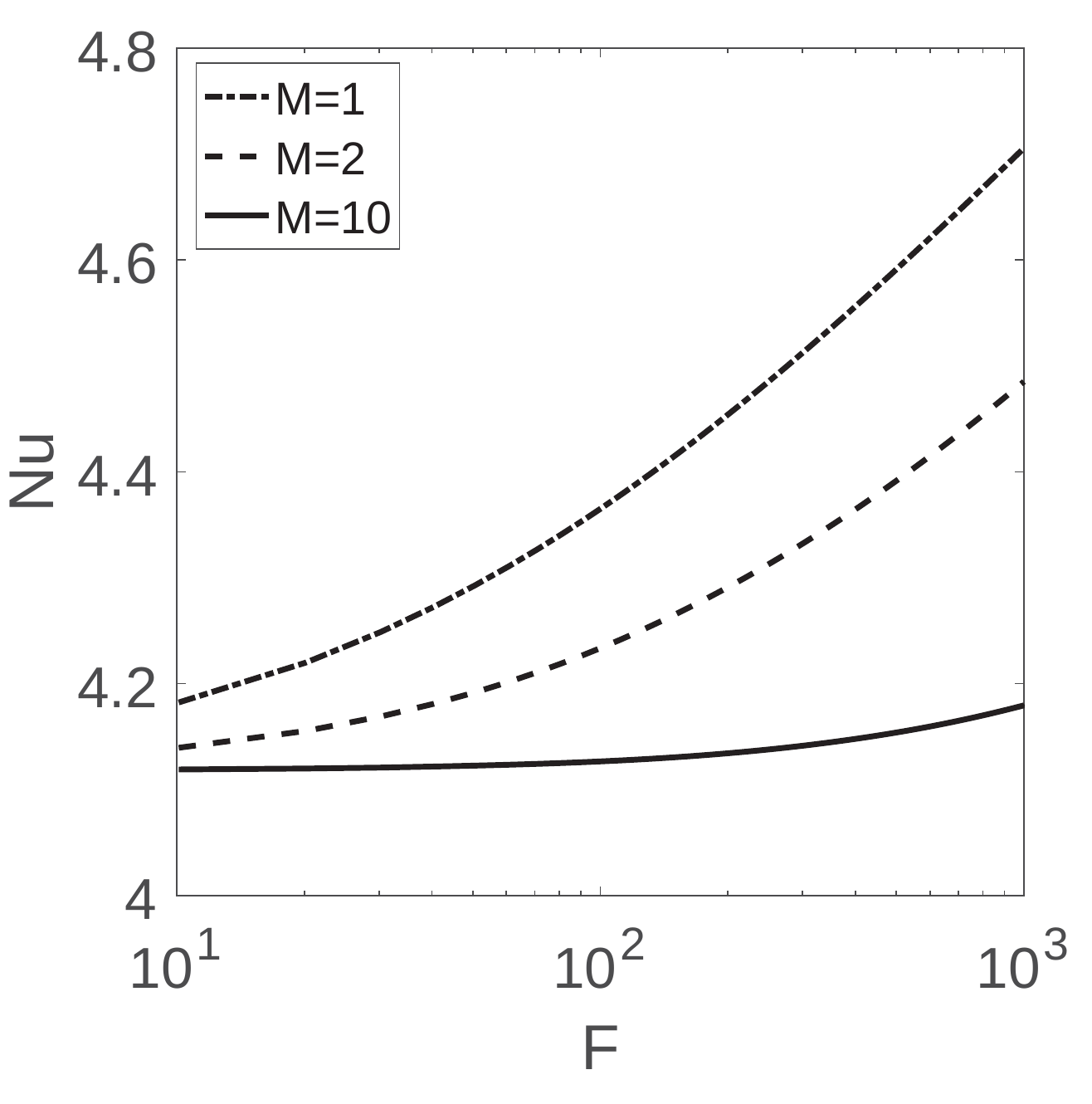}}
\subfigure[]{\label{Fig_2*}\includegraphics[height= 5.8 cm, width= 6.8 cm]{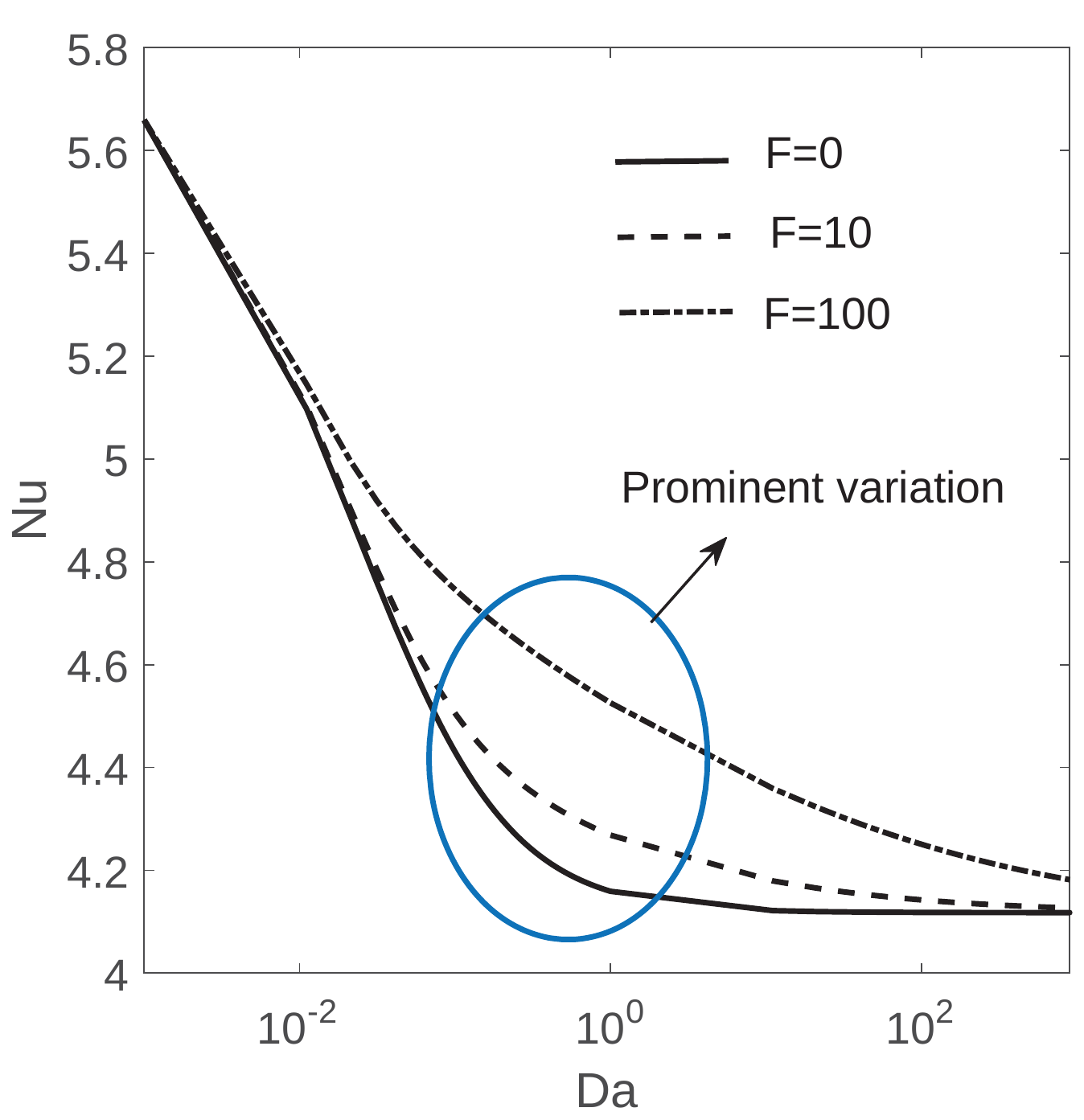}}
\caption{(a) Effect of Forchheimer number, $F$, on the Nusselt number for different viscosity ratio, $M$, for $Da=10^{-2}$, $K=1$.; (b) Nusselt number variation with $Da$ for different inertia parameter, $F$, for $K=1$, $M=1$}
\end{figure}

%\begin{figure}[h]
%\centering
%\subfigure[]{\label{Fig_13}\includegraphics[height= 6 cm, width= 7 cm]{temp_diff_Da_without_dissipation}}
%\subfigure[]{\label{Fig_14}\includegraphics[height= 6 cm, width= 7 cm]{temp_diff_F_without_dissipation}}
%\caption{(a) Temperature profile for various Darcy number for $K=1$, $F=1$, $M=1$; (b) Temperature profile for different Forchheimer number for $Da=1$, $K=1$, $M=1$. }
%\end{figure}

Figure \ref{Fig_6_all} shows $Nu$ as a function of inertia parameter $F$ for different $K$, $\phi$ and $M$. We see that as the inertia parameter $F$ increases, $Nu$ increases. From the physical perspective, the quadratic term that appears in the momentum equation is due to large filtration velocities. Correspondingly, the form drag becomes significant due to comparable friction by the solid obstacles (matrix) inside the porous medium. Hence, increasing inertia parameter leads to hydrodynamic dispersion and hence causes a uniform velocity profile near the center of the channel. When one has a uniform slug like velocity profile across the channel, the term $\left(T_{w}-T_{m}\right)$ decreases and as a result $Nu$ increases. Readers are requested to go through the work by Vafai \& Kim \cite{Vafai1989}, Nield et al. \cite{Nield1996} for a comprehensive discussion. Figure \ref{Fig_10} shows the Nusselt number variation as a function of the inertia parameter $F$ for different anisotropic ratio. We see a significant change in $Nu$ with $K$. For a fixed $Da=10^{-2}$, $\phi=0$ and $F=10^{2}$ the values of $Nu$ are $5.009$, $5.171$, $5.341$ for $K=0.5$, $K=1$, $K=2$, respectively. In accordance with this we see that $Nu$ decreases $16\%$ for $K=0.5$ as that of isotropic situation, $K=1$, and increases $17\%$ for $K=2$ as that of isotropic situation. Hence, for a moderate value of $Da$, anisotropic permeability ratio shows a significant effect on heat transfer coefficient. However, for large values of $Da$, though the inertia effect is expected to be more prominent, the anisotropic effect becomes less prominent, which is analogous to the fact that velocity and temperature tend to a plane-Poiseuille flow situation for large $Da$. We will discuss these asymptotic cases in more detail in the forthcoming section. Figure \ref{Fig_11} depicts the $Nu$ variation with $F$ for different anisotropic angle $\phi$. We see that for a fixed $Da=10^{-2}$, $K=0.2$ and $F=10^{2}$ the values of $Nu$ are $4.841$, $5.045$, $5.171$, for $\phi=0$, $\phi=\pi/4$, $\phi=\pi/2$, respectively. Hence, $Nu$ value is enhanced upto $20\%$ when the anisotropic angle, $\phi$ increases from $0$ to $\pi/4$, and is increased almost $33\%$ when $\phi$ increases from $0$ to $\pi/2$. Hence, heat transfer significantly depends on $K$ and $\phi$.
%===========================================================================
\par
Nusselt number variation as a function of $F$ for viscosity ratio, $M$, is shown in Fig. \ref{Fig_1*}. We see that for low values of $M$ the changes are more prominent and for high values of $M$, $Nu$ changes are very slow. We see that for large values of inertia parameter, $F$, the $Nu$ variation with $M$ is significant. This is because of the higher inertia that enhances the frictional drag, or rather the hydrodynamic dispersion is more for large $F$, and hence the effect of $M$ is prominent. We see a slightly less significant effect for low inertia parameter. When the inertia parameter $F$ is low, say, $F=10$, the values of $Nu$ are $4.182$, $4.139$, $4.119$ for $M=1$, $M=2$, $M=10$, respectively. While, when inertia parameter $F$ is high, say $F=10^{3}$, the values of $Nu$ are $4.706$, $4.485$, $4.179$ for $M=1$, $M=2$, $M=10$, respectively (see Fig. \ref{Fig_1*}). Heat transfer rate decreases only $6\%$ when $F=10$, and viscosity ratio $M$ changes from $1$ to $10$, while the same is almost $52\%$ when $F=10^{3}$. Hence, $M$ has significant effect on the heat transfer rate for large inertial parameter. However, for very large number of $F$, $Nu$ approaches to a fixed asymptotic value $6$ independent of $M$. Such a large viscosity ratio $M=10$ is well inline with the empirical result obtained by Givler and Altobelli \cite{Givler1994}. We have plotted $Nu$ variation with $Da$ for various inertial parameter, $F$ (see Fig. \ref{Fig_2*}). We observe a dramatic impact of $F$ on the heat transfer coefficient, $Nu$. The inertial parameter is prominent in a moderate range of $Da$ (please see the plotted circle in Fig. \ref{Fig_2*}). We see that even though the filtration velocity is high for high $Da$ which includes comparable quadratic drag, in this limit of $Da\rightarrow \infty$, the flow approaches to plane Poiseuille flow situation and approaches to an asymptotic value $6$, independent of $F$.

\section{Conclusions}
We have studied a fully developed forced convection in a fluid saturated anisotropic porous medium bounded by two impermeable wall at a constant heat flux. We have used Brinkman-Forchheimer extended Darcy model in the momentum equation which accounts the inertia of the flow. We have seen that anisotropic permeability ratio and angle has significant effect to the velocity and temperature distribution. Our analysis reveals that for a fixed inertial parameter $F$, the heat transfer rate (Nusselt number) is maximum when the principle axis with higher permeability is parallel to the gravity. We see that increasing the anisotropic ratio enhanced the heat transfer rate, for example, for a fixed value of $Da=10^{-2}$, $F=10^{2}$, $\phi=0$, the Nusselt number increases up to $17\%$ for $K=2$ as that of isotropic situation (i.e., $K=1$) (see Fig. \ref{Fig_10}). Similarly, for a fixed value of $Da=10^{-2}$, $F=10^{2}$, $K=0.2$, the Nusselt number increases up to $33\%$ as $\phi$ changes from $0$ to $\pi/2$ (see Fig. \ref{Fig_11}). We also see that, Nusselt number variation alters significantly with the change of anisotropic ratio and angle. We see that the effect of anisotropy and inertia is prominent in moderate values of $Da$, and approaches to a fixed value $Nu=6$, for small value of $Da$ and to an asymptotic value $4.11$, for large value of $Da$. We have seen the viscosity ratio also effects the flow and heat transfer rate. In view of this we expect that since the heat transfer rate significantly alter, which has certain application in heat exchangers, cold plate devices, the inclusion of the anisotropy in presence of inertia is very useful.

\end{document}